# Lithography defined semiconductor moirés with anomalous in-gap quantum Hall states

Wei Pan[1,*], D. Bruce Burckel[2], Catalin D. Spataru[1], Keshab R. Sapkota[2], Aaron J. Muhowski[2], Samuel D. Hawkins[2], John F. Klem[2], Layla S. Smith[3,1], Doyle A. Temple[3], Zachery A. Enderson[4,5], Zhigang Jiang[5], Komalavalli Thirunavukkuarasu[6,7], Li Xiang[7], Mykhaylo Ozerov[7], Dmitry Smirnov[7], Chang Niu[8], Peide D. Ye[8], Praveen Pai[9], and Fan Zhang[9,*]

[1]Sandia National Laboratories, Livermore, California 94551, USA
[2]Sandia National Laboratories, Albuquerque, New Mexico 87185, USA
[3]Department of Physics, Norfolk State University, Norfolk, Virginia 23504, USA
[4]Oak Ridge Institute for Science and Education Postdoctoral Fellowship, Oak Ridge, Tennessee 37831, USA
[5]School of Physics, Georgia Institute of Technology, Atlanta, Georgia 30332, USA
[6]Department of Physics, Florida A&M University, Tallahassee, Florida 32307, USA
[7]National High Magnetic Field Laboratory, Tallahassee, Florida 32310, USA
[8]School of Electrical and Computer Engineering, Purdue University, West Lafayette, Indiana 47907, USA
[9]Department of Physics, The University of Texas at Dallas, Richardson, TX, 75080, USA

**Abstract:**

Quantum materials and phenomena have attracted great interest for their potential applications in next-generation microelectronics and quantum-information technologies. In one especially interesting class of quantum materials, moiré superlattices (MSL) formed by twisted bilayers of 2D materials, a wide range of novel phenomena are observed. However, there exist daunting challenges such as reproducibility and scalability of utilizing 2D MSLs for microelectronics and quantum technologies due to their exfoliate-tear-stack method. Here, we propose lithography defined semiconductor moirés superlattices, in which three fundamental parameters – electron-electron interaction, spin-orbit coupling, and band topology – are designable. We experimentally investigate quantum transport properties in a moiré specimen made in an InAs quantum well. Strong anomalous in-gap states are observed within the same integer quantum Hall state. Our work opens up new horizons for studying 2D quantum-materials phenomena in semiconductors featuring superior industry-level quality and state-of-the-art technologies, and they may potentially enable new quantum information and microelectronics technologies.

Keywords: lithography defined moirés, compound semiconductor, flat bands, quantum Hall effect, in-gap states, artificial quantum materials

Quantum materials are exciting classes of materials in which quantum phenomena emerge at macroscale without classical counterparts [1]. These materials and phenomena have attracted a great deal of interest for their potential applications in next-generation microelectronics and quantum-information technologies. In one especially interesting class, two same or similar atomically thin layers of 2D van der Waals (vdW) materials such as graphene or transition metal dichalcogenides (TMDs) are overlaid with a small twist angle or lattice mismatch, introducing a moiré periodicity as a new length scale and a moiré band structure as a new basis, not only for energy quantization but also for quantum geometry [2-8]. The so-formed moiré superlattices (MSLs) have become fertile ground for exploring otherwise impossible quantum phenomena, such as Hofstadter butterfly spectrum [9,10] and zero-magnetic-field fractional quantum anomalous Hall effect [11-13]. In particular, when twist bilayer graphene is engineered at its "magic" twist angle, the low-energy moiré bands are nearly flat [14,15], indicating that electrons are heavy and slow, with unusually strong electron-electron interactions. Indeed, a wide range of novel quantum phenomena have been observed, such as quantum geometric superconductivity [16-18] and orbital ferromagnetism [19,20].

The established moiré quantum materials and phenomena are truly exciting. However, vdW MSLs are mainly fabricated using mechanical exfoliation followed by tear-and-stack methods. Thus, MSLs are not only challenging to reproduce because of the inevitable lattice relaxation and spatial inhomogeneity [3, 21] but also incompatible with the state-of-the-art semiconductor processing technologies that exclusively address scalability. In this article, we propose realizing lithography-defined MSLs [22] composed entirely of sophisticated bulk semiconductors (particularly III-V compound semiconductors such as InAs). In these semiconductor MSLs, our theoretical simulations show the formation of flat bands. Like in magic-angle twisted bilayer graphene or TMD moirés, these flat bands are expected to lead to heavy electrons with low kinetic energy and strong electron-electron interactions, giving rise to new strongly correlated electrons.

Figure 1 illustrates our general scheme of creating MSLs in a 2D electron gas (2DEG) formed in a quantum well (QW). As shown schematically in Fig. 1a, we first pattern an in-plane triangular hole (white dots) array. Once the 2DEG in the hole region is etched off, the unetched area forms a honeycomb lattice, namely, artificial graphene (AG) [23-25], as sketched by the red dots in Fig.

1a. Note that the formation of the lateral AG superlattice creates Dirac mini bands with a periodicity-dependent velocity that scales inversely with the superlattice constant [23]. Fig.1b-1d show the AG and MSLs device fabrication in an InAs QW in this work. Interferometric lithography (IL), which utilizes maskless exposure of a photoresist layer with two coherent UV light beams (Fig. 1b), is used to create the AG and MSLs. It provides a flexible, uniform, large-area nano-lithographic capability. To pattern the AG, two separate exposures are used to create the triangular lattice in the resist prior to development. In detail, the interference pattern of two linearly polarized coherent plane waves is used to define the first set of parallel lines in commercially available negative tone photoresist (NR7-500P). Then the sample is rotated by 60 degrees prior to development, and a second exposure is used to define the second set of lines. The exposure dose is controlled so that a triangular array of holes exists after post-exposure bake and development (Fig. 1c). The sample is then etched using reactive ion etching (RIE), and the resist is removed yielding a triangular hole array in the 2DEG. Fig. 1d shows a scanning-electron-microscope (SEM) image of one such AG device.

Two identical AG (AG1 and AG2) with an in-plane angular twist can also be patterned in the same 2DEG. An example of 5 degrees is shown schematically in Fig. 1e. As such, there emerges a much larger hexagonal lattice structure, i.e., a MSL indicated by the yellow hexagon. To create the MSL using IL, two separate, rotated AG patterns are created in the photoresist prior to development. By carefully controlling the angles between the four exposures for the AG patterns, the twist angle between AG1 and AG2 can be precisely defined without requiring any post-processing alignment. The first set of parallel lines for AG1 is defined in the resist followed by the first set of parallel lines for AG2 with a twist angle of 5 degree. This is then followed by the second sets of lines for AG1 and AG2, respectively, at a twist angle of 60 degrees with respective to each of the two first sets. The bottom four panels in Fig. 1b show, schematically, the above process. The laser intensity and expose time for each set of parallel lines are the same. After post-exposure bake, photoresist development and RIE, holes are etched through the 2DEG, and the designed MSL is produced. An SEM image of one such MSL device is shown in Fig. 1f.

To verify the formation of flat bands in our artificial semiconductor MSL, now we show our band structure calculations in Fig. 2. The effective-mass theory in conjunction with muffin-tin potentials is used to describe a 2DEG subjected to a periodic external potential (See Methods). We first show the results for an AG in Fig. 2a and 2d. The strength of potential modulation is set to be 10 in units of $\frac{h^2}{2mL^2}$, where $m$ is the effective mass of the 2DEG in a quantum well, $L$ the lattice constant of the AG, and $h$ the Planck constant. Consistent with a previous study [23], mini Dirac cones merge in the superlattice band structure. Taking $m = 0.023m_e$ for InAs and $L = 200$ nm, the Dirac velocity is calculated to be $\sim 5.3 \times 10^4$ m/s; this is $\sim \frac{1}{20}$ of monolayer graphene's Fermi velocity, demonstrating that the energy bands become flatter with AG potential modulation. Figure 2 also includes two overlaid AG patterns with a commensurate twist angle of $\theta = 38.21°$ in Fig. 2b and $\theta = 5.09°$ in Fig. 2c. The strengths of potential modulation in AG1 (green color) and AG2 (light blue) are set to be $V_1 = V_2 = 10$ in units of $\frac{h^2}{2mL^2}$, and the lattice constants are identically $L$. Figure. 2e and 2f show the electronic band structures for these two commensurate MSLs, featuring mini Dirac points at nearly the same energy of the original Dirac point of AG1 in Fig. 2b. At the large twist angle $\theta = 38.21°$, the Dirac velocity of the MSL is only reduced to $\sim 0.8$ of that of AG, yet at the small twist angle $\theta = 5.09°$ close to the value of our experimental device, ultra-flat bands (with Dirac velocity reduced to $\sim 0.1$ of that of AG) separated from other mini bands emerge. Note that the lattice constant can be made as small as $L = 20$ nm, which is within the capability of extreme-UV lithography. In the Supplementary Materials, we further show that the strength of Rashba spin-orbit coupling in AG can be tuned by simply varying the lattice constant $L$. This may open a new avenue for AG spintronics.

Compared to vdW MSLs, semiconductor 2DEG-based MSLs add two powerful features: intrinsic spin-orbit coupling (SOC) and band gap/topology, and both are accurately designable and continuously tunable. This offers the unique advantage to controllably combine electron-electron interaction and SOC [26], which may enable disentangling of those striking yet elusive quantum phenomena observed in their uncontrolled combinations in vdW MSLs. For example, superconductivity and correlated insulators with unconventional features have been observed in magic-angle TBG [17]. However, it is not yet clear whether and how spin or/and SOC play important roles in producing, enriching, and enhancing (or suppressing) them. To address this outstanding question, a TMD layer was intentionally added to the already delicate device [27,28]. However, it is challenging to control and determine the strength of the proximity-induced SOC.

Another dimension of such superior design is band topology engineering of type-II semiconductors such as InAs/GaSb by varying the well thicknesses [29] or applying an electric field [30], in which Quantum spin Hall insulator [31] and excitonic insulator [32,33] have been achieved.

However, the lattice constant $L$ of AG produced by the UV IL process is limited to ~ 200 nm. For a 1-degree MSL, the MSL constant ($L_m$) is ~ 10 µm and the carrier density hosted in the first mini band, $n_s = 8/(\sqrt{3}L_m^2)$ [17], is ~$10^6$ cm$^{-2}$. To raise $n_s$ to a practical value, one can either increase the twist angle to ~10 degrees or lower $L$ to the 20 nm range by employing e-beam lithography, focused-ion beam milling [34], or even EUV IL. This boosts $n_s$ to ~$10^{11}$ cm$^{-2}$ accessible to InAs quantum wells. Note that the magic angle of 2DEG-based MSL if existing should be much larger than that of twisted bilayer graphene, because of a much smaller Dirac velocity and much stronger "interlayer" tunneling (as two AGs are defined in the same 2DEG).

In the following, we will present our experimental results of magneto transport studies in a specimen of IL-fabricated 2DEG-based MSLs. Similar results have also been observed in other specimens cut from the same wafer. InAs QWs are chosen as the starting material for the following reason. It is known that the Fermi level of InAs normally pins above the bottom of conduction band, this helps avoid the depletion length issue as encountered in GaAs QWs [35]. Figure 3a shows the longitudinal resistance $R_{xx}$ and Hall resistance $R_{xy}$ taken in a non-patterned 2DEG in an InAs QW. At low $B$ fields, $R_{xx}$ displays the SdH oscillations, and $R_{xy}$ is linear with the $B$ field. An electron density of ~ 6.8×$10^{11}$ cm$^{-2}$ and mobility of ~ 1×$10^6$ cm$^2$/Vs are deduced for this InAs quantum well. At high $B$ fields, well-documented QHE is observed – $R_{xx}$ is vanishingly small whereas $R_{xy}$ is quantized to the value of $h/\nu e^2$, where $\nu$ is the Landau level filling factor, $h$ the Planck constant, and $e$ the electron charge. The observation of well-developed QHE attests to the high quality of the starting InAs QW. Figure 3b presents the magneto-transport data in a MSL specimen. The lattice constant for triangular lattice is 250 nm, and the twist angle is 5°. At low $B$ fields, $R_{xx}$ and $R_{xy}$ largely resemble those in the non-patterned sample. A carrier density ~ 7.4×$10^{11}$ cm$^{-2}$ can be deduced. At high $B$ fields, however, transport features in $R_{xx}$ become very different from the non-patterned counterpart. Particularly, in the $\nu = 5$ quantum Hall state, strong in-gap states are observed. Moreover, the strong in-gap states in our MSL device only appears in the $R_{xx}$ trace, and there is no visible change in $R_{xy}$. In fact, $R_{xy}$ remains quantized to the expected value of $h/5e^2$. This is anomalous, given that in the quantum Hall regime a dissipative non-zero $R_{xx}$ always accompanies a non-quantized Hall resistance. As we shall analyze below using the Diophantine equation, our system realizes an extreme limit that has never been achieved by graphene and TMD moirés. In Fig. 3c, for $R_{xx}$ and $R_{xy}$, the two traces for the $B$ field sweeping up and down overlap with each other perfectly. This demonstrates the reproducibility of the in-gap states and that they are not artifacts.

Now we use the Diophantine equation, $n/n_s = t \times \phi/\phi_0 + s$ [36,37], which has been widely used to analyze the magneto transport data of moiré systems [3,38-41], to analyze the strong in-gap states. Here $t$ is the integer (or fractional) value of the quantized Hall conductance, $s$ is the filling of mini energy bands for a periodically modulated 2DEG, $n_s = 1/A$ with $A$ the area of superlattice unit cell, $\phi = B \times A$ the magnetic flux, and $\phi_0 = h/e$ the magnetic flux quantum. For the sake of data analysis, we rewrite this equation as $n = t \times B/\phi_0 + s/A$. Then, it follows that each $R_{xx}$ minima corresponds to an integer value of $s$, with the constraint that all their resulting $(n \times \phi_0 - t \times B)/s$ values should yield the same constant $\phi_0/A$. Table 1 lists the corresponding magnetic field $B$ and the fitted parameter values of ($t$, $s$) for each in-gap $R_{xx}$ minima. In Fig. 3d we plot the value of $(n \times \phi_0 - t \times B)/s$ as a function of $B$, and all the data points indeed fall onto a line of a constant value of ~ 0.15, from which we deduce $A$ ~ 28 ×10$^{-15}$ m$^2$. Remarkably, while this area is much smaller than the area of MSL unit cell (~ 8.1×10$^{-12}$ m$^2$), it is surprisingly close to the area of the triangular AG unit cell shown in Fig. 1c.

A couple of remarks are in order. First, in principle, the Diophantine equation should include both the AG and MSL potential modulations as follows: $n = t \times B/\phi_0 + s_1/A_1 + s_2/A_2$, where $A_1$ and $A_2$ are the areas of AG and MSL unit cells, respectively. Since $A_2 = 8.1 \times 10^{-12}$ m$^2$ is much larger than $A_1 = 2.8 \times 10^{-14}$ m$^2$, for a practical $s_2$, $s_2/A_2$ is negligibly small. Thus, the Diophantine equation is dominated by the AG physics. Second, for $t = 5$, $A_2 = 8.1 \times 10^{-12}$ m$^2$, and a fixed $s_1$, the change in $B$ needed for a change in $s_2$ by 1 is ~ 0.1 mT. Given the specifics of InAs, $\delta B = 0.1$ mT corresponds to an energy variation of 1 mK (here we take $B = 6.5$ T), much smaller than the sample temperature ~ 300mK in our cryogenic system. Thus, to observe any change in $s_2$, one needs to reduce $A_2$ by either increasing the twist angle or to reduce the AG lattice constant from 250 nm to, e.g., 20 nm.

Table 1. The corresponding magnetic field $B$ and the fitted parameter values of ($t$, $s$) for each in-gap $R_{xx}$ minima based on the Diophantine equation $(n \times \phi_0 - t \times B)/s = \phi_0/A$.

| B field | t | s | (n×h/e-t×B)/s |
|---|---|---|---|
| 5.99 | 5 | 6 | 0.145 |
| 6.05 | 5 | 4 | 0.147 |
| 6.26 | 5 | -3 | 0.146 |
| 6.38 | 5 | -8 | 0.132 |
| 6.62 | 5 | -16 | 0.140 |
| 7.08 | 4 | 18 | 0.139 |
| 7.20 | 4 | 14 | 0.146 |
| 7.28 | 4 | 12 | 0.144 |

These strong in-gap states must originate from the moiré engineering. Indeed, for comparison, only much weaker $R_{xx}$ minima are observed in the QHE regime of an AG sample made with the same InAs QW (see Supplementary Materials and Ref. [42]). Importantly, the same quantum Hall state is observed at multiple mini band fillings at a fixed density but a varying $B$ field. As a previously unexplored regime, this is in sharp contrast to the graphene and TMD moiré systems in which multiple quantum Hall or Chern insulator states are observed at the same mini band filling [38-41]. While the graphene and TMD moirés generally exhibit $B/\phi_0 \ll s/A$, our experimental system realizes the opposite limit, $B/\phi_0 \gg s/A$, for the first time, because of the large super-cell area $A$. As we vary the magnetic field $B$, the mini band index $s$ swipes through a series of integers, while the Chern number $t$ remains unchanged. Additionally, in our 2DEG-based MSL, the two AGs are defined in the same 2DEG plane and thus strongly coupled. This is also different from vdW moiré systems in which the interlayer couplings are weak.

To further characterize these in-gap states, we carry out temperature dependent studies of $R_{xx}$ of the strong in-gap states within the same $\nu = 5$ quantum Hall state, as shown in Fig. 4a. Clearly, thermally activated behavior is observed for each in-gap state, and the $R_{xx}$ minima increases with temperature. In Fig. 4b-4e, an Arrhenius plot is displayed for the in-gap states at $B = 6.05$, 6.26, 6.38, and 6.62 T. The red lines are linear fits. It is apparent that for each in-gap state there exist two gaps, one for high temperatures ($\Delta_h$) and the other for low temperatures ($\Delta_l$). In Fig. 4f, the energy gaps for these in-gap states are plotted as a function of $B$. As aforementioned, our system is in the previously unexplored limit, $B/\phi_0 \gg s_1/A_1 \gg s_2/A_2$, because of the large AG unit-cell area $A_1$ and the even larger MSL unit-cell area $A_2$. Thus, as we vary the magnetic field $B$, the mini band index $s_1$ and $s_2$ swipe through series of integers, while the Chern number $t$ remains unchanged. For this reason, we assign the larger gap $\Delta_h$ to the relatively strong AG gap and the smaller gap $\Delta_l$ to the relatively weak MSL gap.

In closing, we note that it was generally believed that the Hofstadter butterfly spectrum can only be observable in the limit of $\phi/\phi_0 \sim 1$ [36]. Surprisingly, the results in our MSL specimen show that the butterfly spectrum is still observable when $\phi/\phi_0 \sim 20$. This new limit may have important implications for studying new many-body states emergent in 2DEG-based MSLs. Lastly, semiconductors such as GaAs and InAs have been the "go-to" materials over the past five decades for classical information technology and may continue their primacy in the quantum era. By engineering their heterostructures into an artificial quantum materials platform, future applications can take the advantage of state-of-the-art semiconductor synthesis and processing such as: 1) crystalline perfection and purity; 2) advanced fabrication techniques; 3) precise reproducibility of AG and MSL without the notorious twist-angle disorder; and 4) compositional control via epitaxy of III-V materials, including heavy (In, Sb) and light (Ga, As) elements for tuning the strength of spin-orbit couplings and vertical engineering for tailoring band topology.

**Methods**

**Electronic transport measurements.** A specimen of 4 mm × 4 mm is cleaved from a MSL patterned wafer and indium contacts are placed along the edge and corner of the specimen. Low frequency (~ 11 Hz) lock-in amplifier techniques with an excitation current of ~ 10 nA is used to measure the transport coefficients $R_{xx}$ and $R_{xy}$. All data are taken in pumped 3He systems and devices are immersed in liquid 3He.

**Band structure calculations.** The AG band structures displayed in Fig. 2 were calculated using the model of muffin-tin potentials [19]. As shown in Fig. 1a, the triangular lattice of white holes act as barriers for electrons and thus produces a honeycomb lattice formed by the red dots. For the single layer AG, we set the strength of the potential given by the holes to $V = 10 \frac{h^2}{2m^*L^2}$ and their radii to $R = \frac{1}{9}L$, and Fourier transformed the muffin-tin potential to the momentum space. Then the Bloch Hamiltonian was diagonalized over a basis of plane wave states, with an energy cutoff $E_c = \left(\frac{200}{3}\right)\frac{h^2}{m^*L^2} = \frac{40}{3}V$ to ensure the convergence of the low-energy band structure. When necessary, an effective mass of $m^* = 0.023 m_e$ (for InAs) and a lattice constant $L = 200$ nm (for IL) were adopted. The commensurate MSL features two identical AG, with the same parameters $V$ and $R$, but rotated at a twist angle $\theta$ without any relative translation. For the MSL calculations, the same energy cutoff $E_c$ was used.

**Acknowledgment**

We are grateful to Jeff Tsao at Sandia for his many insightful comments and suggestions during the wiring of this manuscript. Work at Sandia was supported by the LDRD program. W.P. and K.T. acknowledge support from the DOE, Grant No. DE-SC0024486. W.P. also acknowledges support from the DOE/BES Microelectronics Science Research Center. Z.J., L.X., and D.S acknowledge support from the DOE, Grant No. DE-FG02-07ER46451. P.P. and F.Z. were supported by NSF under grants DMR-1945351, DMR-2324033, and DMR-2414726; they acknowledge the Texas Advanced Computing Center (TACC) for providing resources that have contributed to the research results reported in this work. Part of the measurements were carried out at the National High Magnetic Field Laboratory, which is supported by the NSF Cooperative Agreement (No. DMR-2128556) and the State of Florida. Part of device fabrication was performed at the Center for Integrated Nanotechnologies, a U.S. DOE, Office of BES, user facility. Sandia National Laboratories is a multimission laboratory managed and operated by National Technology and Engineering Solutions of Sandia LLC, a wholly owned subsidiary of Honeywell International Inc. for the U.S. DOE's National Nuclear Security Administration under contract DE-NA0003525. This written work is authored by an employee of NTESS. The employee, not NTESS, owns the right, title and interest in and to the written work and is responsible for its contents. Any subjective views or opinions that might be expressed in the written work do not necessarily represent the views of the U.S. Government. The publisher acknowledges that the U.S. Government retains a non-exclusive, paid-up, irrevocable, world-wide license to publish or reproduce the published form of this written work or allow others to do so, for U.S. Government purposes. The DOE will provide public access to results of federally sponsored research in accordance with the DOE Public Access Plan.

**Supporting Information**

I. Electronic transport in an artificial graphene sample and the Diophantine analysis

II. Tunable strong spin-orbit coupling in AG

**Authors contributions**

W.P., D.B.B., C.D.S., J.F.K., and F.Z. conceived the project. A.J.M., S.D.H., and J.F.K. contributed to materials design and growth; W.P., D.B.B., K.S.S., C.N. contributed to device fabrication; W.P., L.S.L.S., Z.E., K.T., L.X., M.O., and D.S. contributed to characterization and measurements; C.D.S. and P.P. contributed to theoretical simulations. D.A.T. supervised the work at Norfolk, Z.J. supervised the work at Ga Tech, P.D.Y. supervised the work at Purdue, and F.Z. supervised the work at UT Dallas. W.P. and F.Z. wrote the manuscript with inputs from all authors.

**Competing interests**

The authors declare no competing interests.

**Figures**

**Figure1**

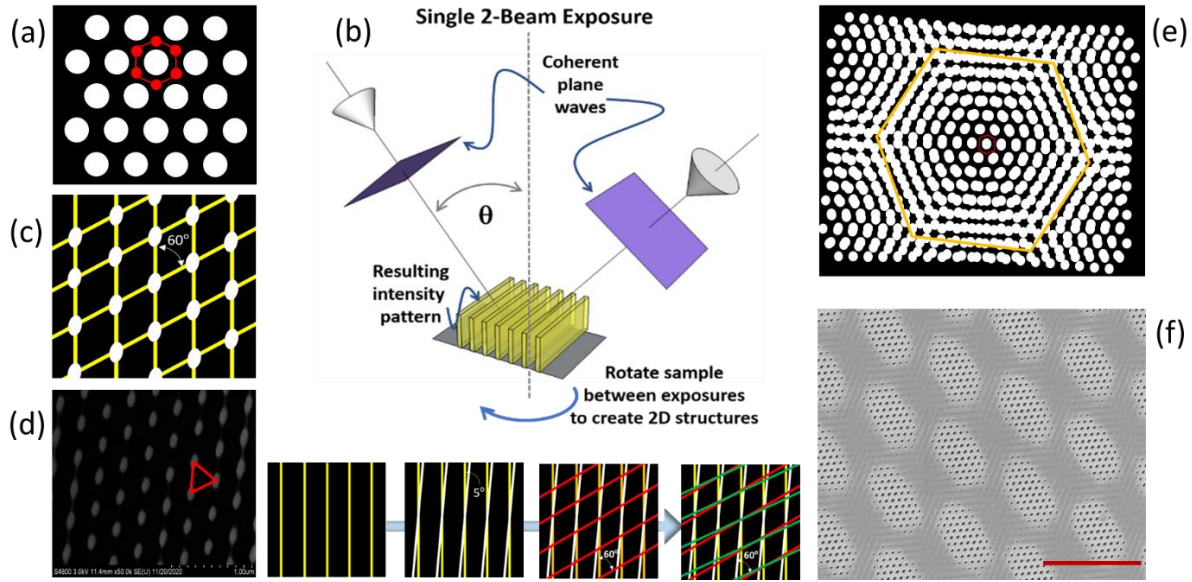

Figure 1: **General scheme of creating moiré superlattices in a 2D electron gas (2DEG) formed in a semiconductor quantum well (QW)**. (a) Schematic representation of artificial graphene (AG). The white dots represent holes etched through the 2DEG (black square), leaving the red areas forming a honeycomb lattice. (b) Schematic depiction of two-beam interferometric lithography (IL), where parallel lines of alternating intensity form due to the interference of two linearly polarized coherent plane waves. The bottom four panels show schematically how a lithography defined moiré superlattice is fabricated using the IL method. (c) Schematic of two sets of parallel lines of interference pattern with a twist angle of 60 degrees. By adjusting the laser intensity and expose time, only the photoresist at the crossing points (brown dots) is fully exposed. (d) SEM image of a triangular lattice of etched holes through a 2DEG in an InAs QW. (e) Schematic of two AG patterns with a twist angle of 5°. The large supercell is indicated by the yellow hexagon. (f) SEM image of a moiré superlattice in an InAs QW. The scale bar is 5 μm.

**Figure 2**

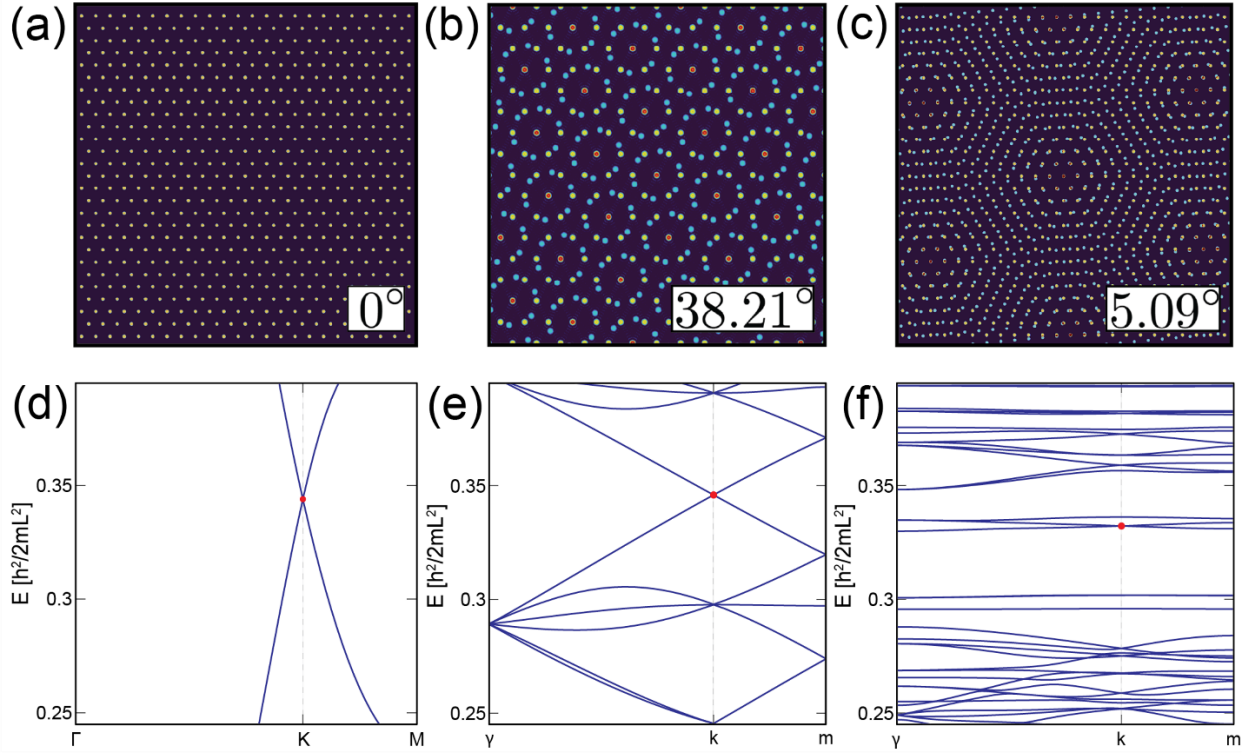

Figure 2: **Plots of the real-space potential profiles and resulting band structures of the artificial graphene (AG) and moiré superlattice (MSL) systems**. For all potential holes, $V = 10\frac{h^2}{2mL^2}$ and $\frac{R}{L} = \frac{1}{9}$. (a) Plot of the muffin-tin potential profile that produces the AG. (b) By rotating a second set of the same AG by $\theta = 38.21°$, a commensurate MSL pattern appears. The yellow and blue potential holes indicate each of the two layers, and the red holes are where they overlap. (c) The same as (b) but for a commensurate MSL pattern with $\theta = 5.09°$. (d)–(f) The calculated band structures of (a)–(c), respectively. The red dots indicate the Dirac points.

**Figure 3**

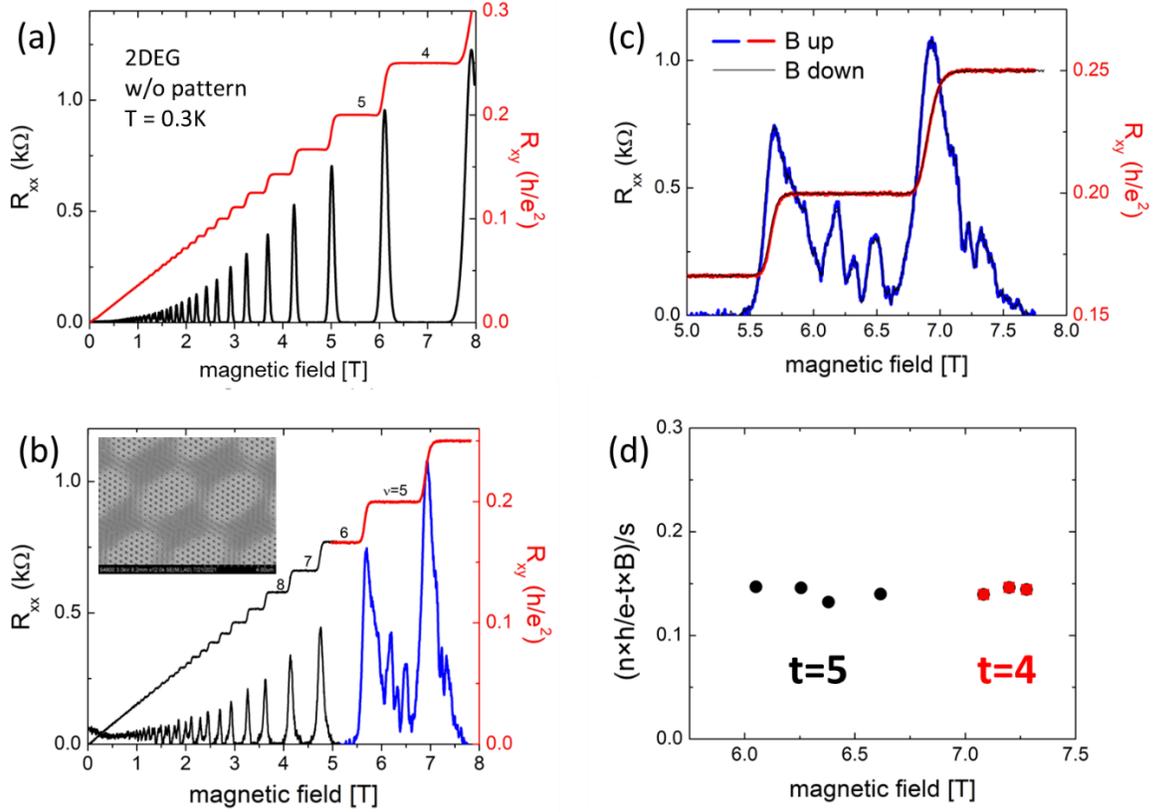

Figure 3: **Anomalous in-gap states in lithography defined semiconductor moiré**. (a) Electronic transport coefficients, magnetoresistance $R_{xx}$ and Hall resistance $R_{xy}$, of the 2DEG in an InAs quantum well. (b) $R_{xx}$ and $R_{xy}$ of a 2DEG-based moiré fabricated with the same QW as (a). The blue part highlights the anomalous in-gap states, in the quantum Hall state at Landau level filling $\nu = 5$. The quantized Hal resistance is unaffected by these in-gap states. (c) The $R_{xx}$ (and $R_{xy}$) traces for the magnetic field sweeping up and down. The two traces overlap perfectly, demonstrating the robustness of the in-gap states. (d) Diophantine analysis of the in-gap states. All the data points show similar values of ~ 0.15.

**Figure 4**

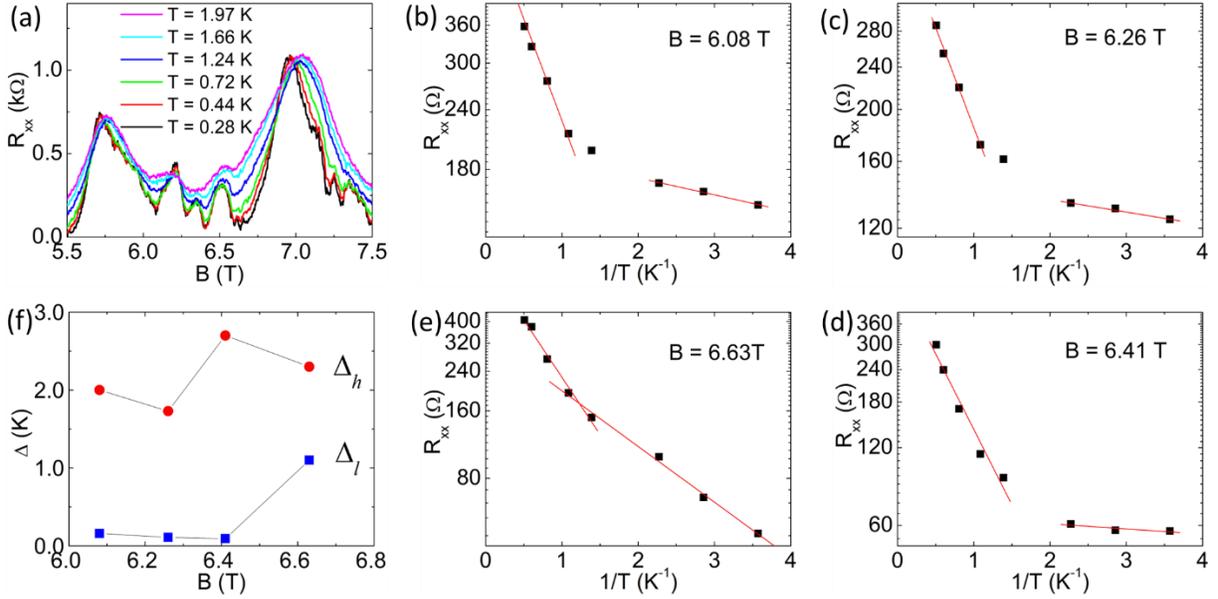

Figure 4: **Energy gaps of the anomalous in-gap states**. (a) Temperature dependence of $R_{xx}$ for the in-gap states in the $\nu = 5$ quantum Hall state. (b)-(e) Arrhenius plots for the four in-gap states at the magnetic fields $B = 6.08$, 6.26, 6.41, and 6.63 T, respectively. It is clearly seen that there exist two slopes for each case, signaling a two-gap nature for the in-gap states. (f) Energy gaps plotted as a function of the magnetic field.

**TOC graphic**

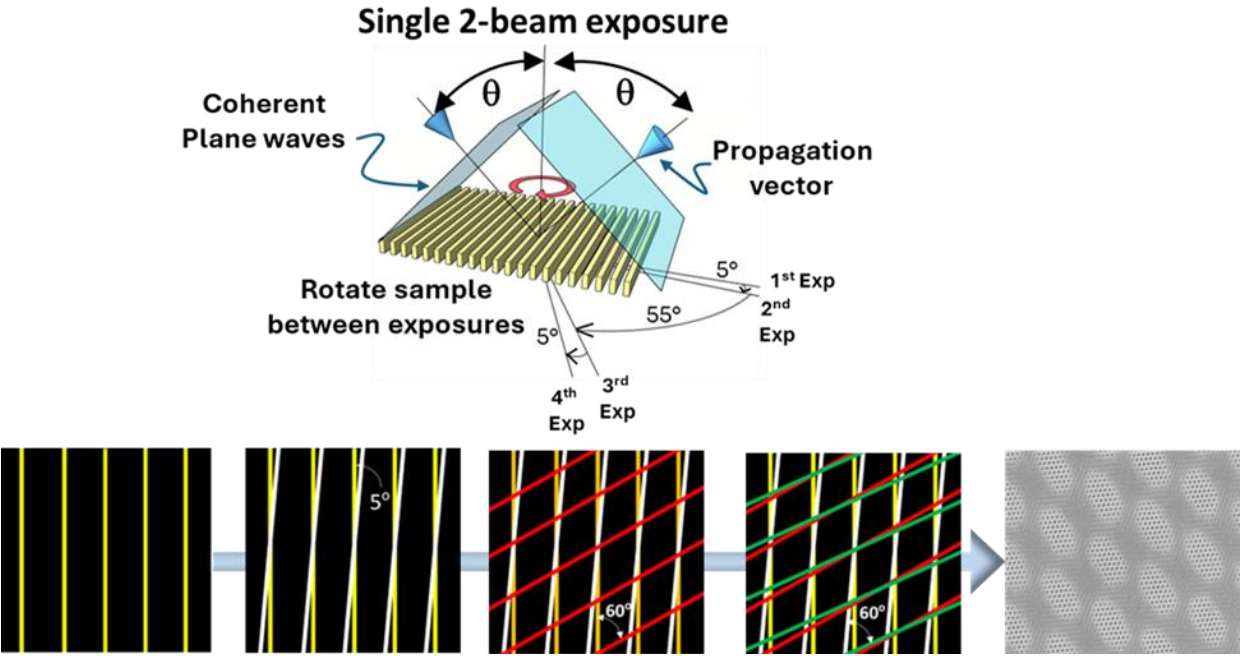

**Supporting Information**

## I. Electronic transport in an artificial graphene sample and the Diophantine analysis

In Fig. S1a, we show the $R_{xx}$ and $R_{xy}$ data taken in an AG sample fabricated in a specimen cut from the same InAs QW wafer as in Fig. 3a. The lattice constant of the triangular arrays is 250 nm. The same plot has also been used in a recent publication [42]. Overall, the $R_{xx}$ and $R_{xy}$ traces in this AG sample resemble those in the unpatterned sample though the density is higher, ~ $8\times10^{11}$ cm$^{-2}$. In the high B field regime, again, quantized Hall states are formed. However, careful examinations of $R_{xx}$ shows additional minima, as marked by the vertical red lines (Fig. S1b), inside the gapped quantum Hall states (the magnetic field positions of the integer quantum Hall states are marked by short black lines). We further analyze these additional $R_{xx}$ minima inside the quantum Hall state, using the Diophantine equation. To perform this scaling analysis, we assign an integer value of s to each in-gap minimum, with the constraint that $(n\phi_0 - tB)/s$ would be roughly constant, since it is directly determined by $A$, the area of superlattice unit cell.

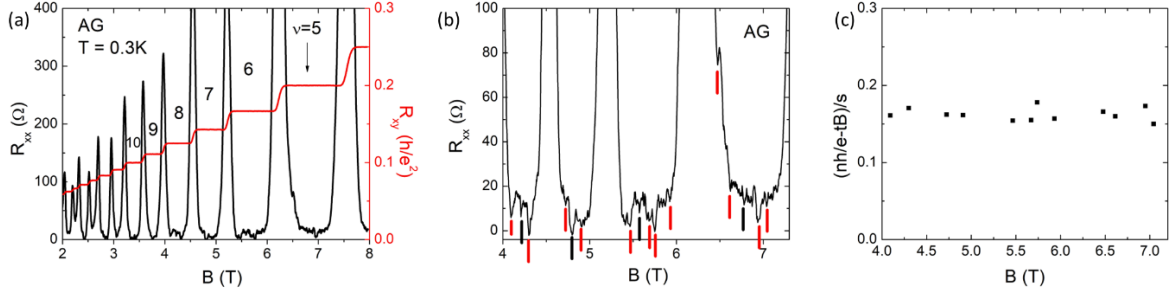

Figure S1. (a) $R_{xx}$ and $R_{xy}$ data taken in an AG sample fabricated in a specimen cut from the same InAs QW wafer as in Fig. 3a. The lattice constant of the triangular arrays is 250 nm. (b) Zoom-in of (a) from 4 T to 7.5 T. (c) Plot of the value of $(n\phi_0 - tB)/s$ as a function of $B$.

Table S1 lists the parameters ($B$, $t$, and $s$) for each in-gap $R_{xx}$ minima. In Figure S1c, we plot the value of $(n\phi_0 - tB)/s$ as a function of $B$. It is remarkable that all the data points are roughly equal, with an average of ~ 0.16. From this constant value, we can deuce the unit area of superlattices, and $A \approx 25\times10^{-15}$ m$^2$, close to the triangular lattice area (red color) in the AG sample, as shown in Fig. 1c.

| B field | t | s | (n×h/e-t×B)/s |
|---|---|---|---|
| 4.093 | 8 | 6 | 0.160 |
| 4.299 | 8 | -4 | 0.170 |
| 4.723 | 7 | 4 | 0.162 |
| 4.908 | 7 | -4 | 0.161 |
| 5.464 | 6 | 6 | 0.154 |
| 5.670 | 6 | -2 | 0.155 |
| 5.737 | 6 | -4 | 0.178 |
| 5.932 | 6 | -12 | 0.157 |
| 6.476 | 5 | 8 | 0.166 |
| 6.614 | 5 | 4 | 0.160 |
| 6.950 | 5 | -6 | 0.173 |
| 7.042 | 5 | -10 | 0.150 |

## II. Tunable strong spin-orbit coupling in AG

In the main text, we have set the spin-orbit coupling (SOC) strength to be zero for simplicity. However, this can be added to our calculations in a way that is experimentally feasible. Take Rashba SOC for example here. In the original electron gas system, Rashba SOC is described by $H_{SOC} = \alpha\,(\vec{k} \times \vec{s}) \cdot \hat{z}$ and $\alpha = 5 \times 10^{-12}$ eV·m is a realistic number for the material. Near the Dirac point of graphene or AG, the Rashba SOC is described by $H_{SOC} = \lambda_R\,(\sigma_x s_y - \sigma_y s_x)$, producing two splittings of $2\lambda_R$ in band energy. To a good approximation, for AG $\lambda_R \propto \alpha|\Gamma K| \propto \alpha/L$, where $L$ is the designable AG lattice constant. As calculated and plotted in Fig. S2, $\lambda_R$ increases from 0.05 meV to 0.2 meV as $L$ is decreased from 200 nm to 50 nm. (*Note that the SOC strength is in the order of 0.001 meV in real graphene.*) This confirms the tunability of strong SOC in our artificial structures.

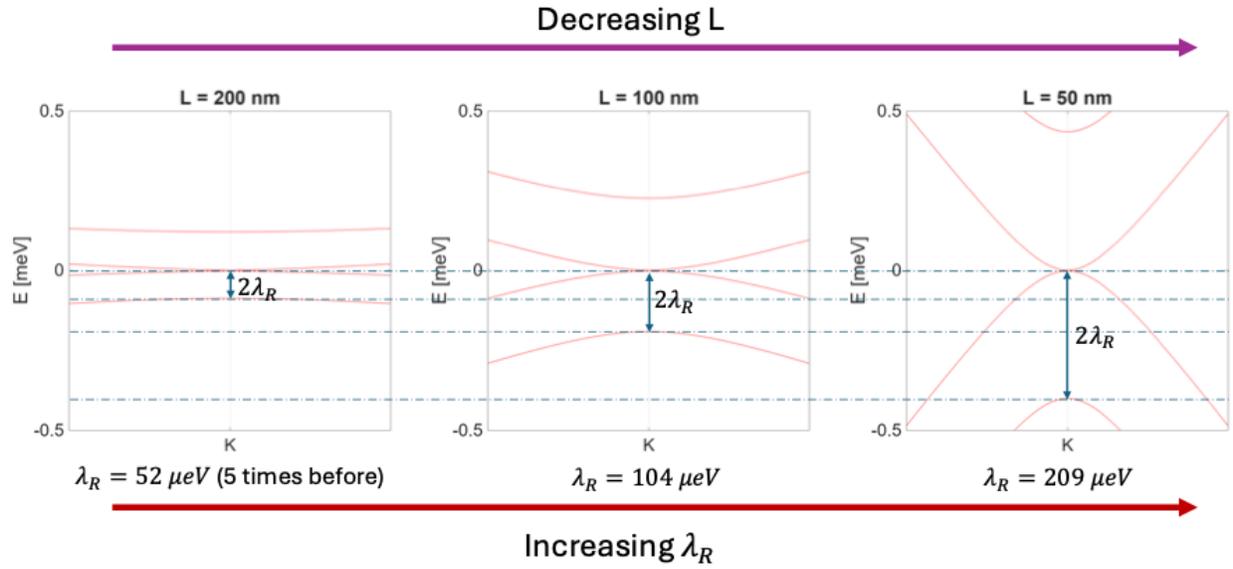

Figure S2. The calculated electronic band structures of artificial graphene near one Dirac point. From left to right, the engineered lattice constants are $L$ = 200 nm, 100 nm, and 50 nm, and the fitted Rashba SOC strengths are $\lambda_R$ = 0.052 meV, 0.104 meV, and 0.209 meV.